\documentclass[aps,prl,twocolumn,groupedaddress]{revtex4}
\usepackage{amsmath}
\usepackage{amsfonts}
\usepackage{amssymb}
\usepackage{graphicx}
\usepackage{epsfig}

\begin{document}

\title{Strong Shock Waves and Nonequilibrium Response in a One-dimensional Gas: \\ 
a Boltzmann Equation Approach}

\author{Pablo I. Hurtado}

\affiliation{Department of Physics, Boston University, Boston, MA 02215, USA, and \\
Institute \emph{Carlos I} for Theoretical and Computational Physics, Universidad de Granada, 18071 Granada, Spain.}

\today

\begin{abstract}
We investigate the nonequilibrium behavior of a one-dimensional binary fluid on the basis of Boltzmann equation,
using an infinitely strong shock wave as probe.
Density, velocity and temperature profiles are obtained as a function of the mixture mass ratio $\mu$. 
We show that temperature overshoots near the shock layer, and that heavy particles are denser, slower and cooler than light 
particles in the strong nonequilibrium region around the shock. The shock width $\omega(\mu)$, which characterizes the
size of this region, decreases as $\omega(\mu) \sim \mu^{1/3}$ for $\mu \to 0$. In this limit, two very different length scales control 
the fluid structure, with heavy particles equilibrating much faster than light ones. 
Hydrodynamic fields relax exponentially toward equilibrium, $\phi(x)\sim \textrm{exp}[-x/\lambda]$. 
The scale separation is also apparent here, with two typical 
scales, $\lambda_1$ and $\lambda_2$, such that $\lambda_1\sim \mu^{1/2}$ as $\mu \to 0$, while $\lambda_2$, which is the slow
scale controlling the fluid's asymptotic relaxation, increases to a constant value in this limit. These results are discussed
at the light of recent numerical studies on the nonequilibrium behavior of similar 1d binary fluids.
\end{abstract}



\maketitle

\section{I. Introduction}

When a piston moves at constant velocity into a fluid, it generates a shock wave.\cite{Whitham} This is a strong perturbation 
that drives the system far from equilibrium. Studying the structural properties of the shock wave, and how equilibrium is restored 
behind it, we may extract valuable information on the fluid's transport properties, the damping of nonequilibrium fluctuations, 
etc. That is, we may use the shock wave as a probe to better understand the nonequilibrium response of the fluid.
This is particularly appealing for one-dimensional (1d) systems. Their nonequilibrium behavior has attracted much attention 
during the last years.\cite{review}-\cite{Baowen} This follows from two main reasons. On one hand,  the dimensional constrain 
characteristic of 1d systems plays an essential role in many real structures, ranging from carbon nanotubes\cite{carbon} to 
anisotropic crystals\cite{crys}, magnetic systems\cite{magnet} solid copolymers\cite{poly}, semiconductor wires\cite{wires}, 
zeolites\cite{zeo}, Bose-Einstein condensates\cite{Bose}, colloids in narrow channels\cite{SF}, DNA, nanofluids, etc.\cite{review} 
On the other hand, the simplicity and versatility of 1d models allow one to tackle fundamental questions in 
nonequilibrium statistical mechanics, as those related to irreversibility, normal transport, equipartition, local thermodynamic 
equilibrium, etc.\cite{review} 

In this way, it has been found that the low dimensionality can give rise to anomalous transport properties in 1d fluids. To be 
more specific, the \emph{single-file} (SF) constraint characteristic of 1d fluids introduces strong correlations between 
neighboring particles, which asymptotically suppress mass transport \cite{SF} and enhance 
heat conduction \cite{Fourier1,Fourier2,Narayan,untercio,dosquintos,Boltzmann,uncuarto,Baowen}. 
In particular, it is currently believed (see however \cite{Fourier1,Fourier2}) that 1d fluids with 
momentum-conserving interactions and non-zero total pressure 
exhibit a thermal conductivity $\kappa$ that diverges as $\kappa\sim L^{\gamma}$ 
when the system size $L\to \infty$.\cite{Narayan,untercio,dosquintos,Boltzmann,uncuarto} 
However, there is no complete agreement yet on the exponent $\gamma>0$: 
there exist strong numerical and theoretical evidences pointing out that $\gamma=1/3$ \cite{Narayan,untercio},
although mode-coupling theories and Boltzmann equation analysis predict $\gamma=2/5$ \cite{dosquintos,Boltzmann}, 
and other numerical results are closer to $\gamma=1/4$ \cite{uncuarto}.
Conservation of momentum seems to be crucial;
as soon as this symmetry is broken, normal heat transport is recovered.\cite{Baowen} 
Moreover, momentum-conserving 1d fluids typically exhibit other anomalies when perturbed away from equilibrium, 
as for instance dynamic non-equipartition of energy, emergence of multiple relaxation scales, 
energy localization, etc.\cite{Fourier1,Fourier2,untercio,uncuarto,Jou} In general,
the behavior of 1d fluids far from equilibrium still poses many intriguing questions which require further investigation.

In this paper we probe the nonequilibrium response of a 1d model gas. In particular, we
address the problem of an \emph{infinitely strong} shock wave propagating into a 1d binary fluid, within the 
context of Boltzmann equation.\cite{Resibois,Mazur,Harris}  The shock wave characterizes the transition between two different 
asymptotic equilibrium states of the fluid. For strong shock waves (to be specified latter), 
this transition happens within molecular scales, so kinetic (Boltzmann) equations must be used. In fact, one may think of the shock 
wave problem as the simplest case study dominated by the nonlinearity of Boltzmann equation.\cite{Cercignani}
As we will see below, Boltzmann equation provides a powerful tool to investigate the observed structural and relaxational anomalies 
in 1d fluids. Kinetic theory methods have been applied with success to study strong shock waves in high-dimensional 
\emph{simple} fluids.\cite{Grad,Mott,Cercignani} However, in one dimension, a simple (i.e. mono-component) fluid 
constitutes a pathological limit.\cite{Jepsen} This follows from the fact that
particles with equal masses in 1d interchange their velocities upon collision, so a 1d mono-component fluid can be regarded 
to a large extent as an ideal gas of non-interacting particles.\cite{equalmass} We may restore ergodicity and 
relaxation in velocity space by introducing different masses. The simplest case is that of a binary mixture. In particular, we study in 
this paper a binary 1d fluid composed by two species of hard-point particles, with masses $m_1<m_2$ and \emph{equal} 
concentrations. Energy transport in this model has already been extensively studied via molecular dynamics 
simulations.\cite{Fourier1,Fourier2,untercio,uncuarto,Pablo}

The structure of the paper is as follows. In section II, we write down the relevant Boltzmann equation for our 1d binary fluid, 
and extend a method by H. Grad\cite{Grad,Cercignani} to study the structure of an infinitely strong shock wave propagating 
through the fluid. The results of this approach are described in section III. In particular, we study there the shock hydrodynamic 
profiles, paying special attention to the shock width and the relaxation toward equilibrium, as a function of the mass ratio. Finally, 
in section IV we discuss our results.

\section{II. The Shock-Wave Problem}

The Boltzmann equation for a 1d binary fluid in a reference frame moving with the shock wave 
reads,\cite{Resibois,Mazur,Harris,Cercignani,Maxwell1d}
\begin{eqnarray}
v\frac{\text{d} f_1}{\text{d} x} & = & {\cal Q}_1(f_1,f_2) - f_1 \nu (f_2), \label{eqB1} \\
v\frac{\text{d} f_2}{\text{d} x} & = & {\cal Q}_2(f_2,f_1) - f_2 \nu (f_1).
\label{eqB2}
\end{eqnarray}
Here $f_i(x,v)$ is the probability density for finding a particle of type $i=1,2$ (i.e. mass $m_i$) at position $x$ with velocity $v$, 
and ${\cal Q}_i$ and $\nu$ represent the gain term and collision frequency in the collision operator, respectively. They are defined as,
\begin{eqnarray}
{\cal Q}_i(f_i,f_j) & = & \int_{-\infty}^{+\infty} \text{d}w \ \vert v-w \vert f_i(x,v_i') f_j(x,w_i'), \label{Qdef} \\
\nu(f_i) & = &  \int_{-\infty}^{+\infty} \text{d}w \ \vert v-w \vert f_i(x,w). \label{nudef}
\end{eqnarray}
Index $j$ refers here to particle species other than $i$, and $(v_i',w_i')$ are pre-collisional velocities that after collision 
give rise to velocities $(v,w)$ for the pair $(i,j)$, namely,
\begin{eqnarray}
v_i'(v,w) & = & \frac{(m_i-m_j)v + 2m_jw}{m_i+m_j}, \nonumber \\
w_i'(v,w) & = & \frac{(m_j-m_i)w + 2m_iv}{m_i+m_j}.
\label{prev}
\end{eqnarray}
Notice that eqs. (\ref{eqB1})--(\ref{eqB2}) only include the effect of cross-collisions between the different species, and no 
self-collisions between like particles. 
This is reflecting the one-dimensional character of our model fluid,
and it is a main difference with Boltzmann equation for mixtures in higher dimensions.

The structure of the shock wave can be deduced from eqs. (\ref{eqB1})--(\ref{eqB2}) subject to the boundary conditions,
\begin{equation}
f_i(x,v) \rightarrow G_{i,\pm}(v) \quad \text{as} \quad x\rightarrow \pm \infty ,
\label{bc}
\end{equation}
where
\begin{equation}
G_{i,\pm}(v) \equiv n_{\pm}\sqrt{\frac{m_i}{2\pi T_{\pm}}} 
\text{exp}\Big[-\frac{m_i(v-u_{\pm})^2}{2T_{\pm}}\Big].
\label{maxwellian}
\end{equation}
Here $n_{\pm}$, $T_{\pm}$ and $u_{\pm}$ are respectively the number density, temperature and macroscopic 
velocity at $x\rightarrow \pm \infty$.\cite{equaldens} One usually chooses $u_+>0$, which implies (see below) $u_->0$.
This amounts to represent a flow of the fluid mixture from $-\infty$ (upstream or pre-shock region) to $+\infty$ 
(downstream or after-shock region).

We are particularly interested in the structure of an \emph{infinitely strong} shock wave. In this way we can probe the
response of the 1d binary mixture in the strong nonequilibrium regime, far from the linear response region.\cite{Mazur}
The strength of a shock may be defined in several ways. One of them is based on the ratio of downstream to upstream pressures, 
$r=p_+/p_-$, where $p_{\pm}=n_{\pm}T_{\pm}$. Weak shocks have a ratio $r$ close to unity. In this case, 
the shock thickness is large as compared to the mean free path,  and continuum approximations of the type of
Navier-Stokes (or Euler, Burnett, etc.) may be used to characterize the shock structure.\cite{Whitham} However, 
as $r$ grows the shock thickness becomes comparable to the mean free path, making inappropriate the application 
of hydrodynamic approximations.
In this case kinetic equations, as those in (\ref{eqB1})--(\ref{eqB2}), must be used instead. 
We pay attention in what follows to the limit $r \rightarrow \infty$, or equivalently $T_- \rightarrow 0$. 

The parameters entering the two limiting Maxwellians (\ref{maxwellian}) are related due to conservation of particle number, total 
momentum and total energy. Such conservation laws can be expressed as the constancy of the corresponding fluxes,
which yields,
\begin{eqnarray}
n_+u_+ & = & n_-u_-, \label{RH} \\
n_+\big[(m_1+m_2)u_+^2 + 2 T_+  \big] & = & n_-\big[(m_1+m_2)u_-^2 \big],  \nonumber \\
n_+\big[(m_1+m_2)u_+^3 + 6u_+T_+  \big] & = & n_-\big[(m_1+m_2)u_-^3 \big]. \nonumber
\end{eqnarray}
These are the Rankine--Hugoniot conditions, which relate the downstream and upstream values of the flow fields 
in the mixture.\cite{Whitham} Solving this system for the downstream asymptotic flow fields one finds,\cite{trivialsol}
\begin{equation}
n_+=2n_-, \quad u_+=\frac{u_-}{2}, \quad T_+ = \frac{1}{8}(m_1 + m_2)u_-^2.
\label{final}
\end{equation}
Therefore, the downstream asymptotic density (velocity) is twice (half) the upstream one for $r\rightarrow\infty$. 
It must be noticed that these results come from the conservation laws 
characterizing the two-component fluid, and have nothing to do with the 
hydrodynamic behavior of the mixture. In this way it is not strange that even in non-hydrodynamic cases, as the 
equal mass version of our one-dimensional gas,\cite{equalmass} the results (\ref{final}) hold.

The case of a infinitely strong shock has been studied in detail for mono-component gases in dimensions larger than 
one.\cite{Grad, Cercignani} Extending a successful hypothesis by H. Grad,\cite{Grad} we suggest here that
the problem of a shock wave in our 1d mixture is in general well-defined 
in the limit $r\rightarrow \infty$, and that in this case the distribution functions $f_i$ can be decomposed into a singular part, 
$\hat{f}_i$, proportional to a Dirac $\delta$-function on the upstream velocity $u_-$, and a regular part, $\tilde{f}_i$, 
smoother than the former. This is because for $r\rightarrow \infty$, particles in the upstream region exhibit a well-defined 
fixed thermal velocity when measured on the downstream velocity scale. Hence,
\begin{equation}
f_i = \hat{f}_i + \tilde{f}_i, \quad \text{with} \quad \hat{f}_i(x,v) = \hat{n}_i(x) \delta(v-u_-).
\label{decomp}
\end{equation}
The singular particle densities $\hat{n}_i(x)$ control how important the singular component is at a given coordinate $x$. In 
particular, it is evident that $\hat{n}_i(x) \to n_-$ as $x\to -\infty$, and $\hat{n}_i(x) \to 0$ as $x\to +\infty$. Introducing the 
above decomposition (\ref{decomp}) into eqs. (\ref{eqB1})--(\ref{eqB2}), we may split the original Boltzmann equations in two 
different sets, attending to the singular or regular character of the terms,
\begin{eqnarray}
v\frac{\text{d} \hat{f}_i}{\text{d} x} + \hat{f}_i\nu(\tilde{f}_j) = {\cal Q}_i(\hat{f}_i,\hat{f}_j) - 
\hat{f}_i \nu (\hat{f}_j), \label{singular} \\
v\frac{\text{d} \tilde{f}_i}{\text{d} x} -{\cal Q}_i(\tilde{f}_i,\hat{f}_j) - {\cal Q}_i(\hat{f}_i,\tilde{f}_j) + 
\tilde{f}_i\nu(\hat{f}_j)  =  \nonumber \\
 = {\cal Q}_i(\tilde{f}_i,\tilde{f}_j) - \tilde{f}_i \nu (\tilde{f}_j),
\label{regular}
\end{eqnarray}
with $i=1,2$. Eq. (\ref{regular}) for the regular component $\tilde{f}_i$ looks more complicated than the original
Boltzmann equations (\ref{eqB1})--(\ref{eqB2}). However, as advantage, it lacks the singular behavior characterizing for 
$r\rightarrow \infty$ the whole distribution $f_i$. This allows us to perform a simple local Maxwellian approximation for 
$\tilde{f}_i$ that yields meaningful results. In particular, we now assume,
\begin{equation}
\tilde{f}_i(x,v) = \tilde{n}_i(x)\sqrt{\frac{m_i}{2\pi \tilde{T}_i(x)}}
\text{exp}\Big[-\frac{m_i\big(v-\tilde{u}(x)\big)^2}{2\tilde{T}_i(x)}  \Big]. 
\label{maxw}
\end{equation}
Here $\tilde{n}_i(x)$, $\tilde{u}(x)$ and $\tilde{T}_i(x)$ are the number density, velocity and temperature associated
to $\tilde{f}_i$. Now $\tilde{n}_i(x) \to 0$ as $x \to -\infty$ and $\tilde{n}_i(x) \to n_+$ as $x \to +\infty$. 
We have also assumed at this point that the regular flow velocity $\tilde{u}(x)$ does not depend on particle 
species. This assumption seems natural given the one-dimensional character of the system, which forces neighboring particles 
to move coherently on average. Therefore the problem of the shock wave 
structure reduces to computing the spatial dependence of 7 different flow fields, namely $\hat{n}_i(x)$, $\tilde{n}_i(x)$, 
$\tilde{T}_i(x)$ and $\tilde{u}(x)$, $i=1,2$. 

It is important to notice that the Maxwellian approximation (\ref{maxw}) is not a local thermodynamic equilibrium (LTE) hypothesis,
since it only affects the regular component $\tilde{f}_i(x,v)$ of the distribution function. The singular, delta-like component 
$\hat{f}_i(x,v)$ accounts for the most important nonequilibrium effects. However, it is natural to question the validity of 
(\ref{maxw}) for the present strong nonequilibrium problem. This approximation is just the zeroth-order
term in a formal expansion of $\tilde{f}_i(x,v)$ around local Maxwellian behavior, in the spirit of 
Sonine polynomials expansions, Gram-Charlier and Edgeworth series, etc.\cite{Abramowitz,gauss,wolfram} One would expect 
that higher-order terms in this expansion could be relevant near the shock layer. In fact, such slight deviations have been observed 
in mono-component gases in higher dimensions.\cite{Cercignani} However, these small corrections 
have negligible effects on the shock hydrodynamic profiles,\cite{Cercignani} and therefore are not important for our discussion here.

\begin{figure}[t]
\centerline{
\psfig{file=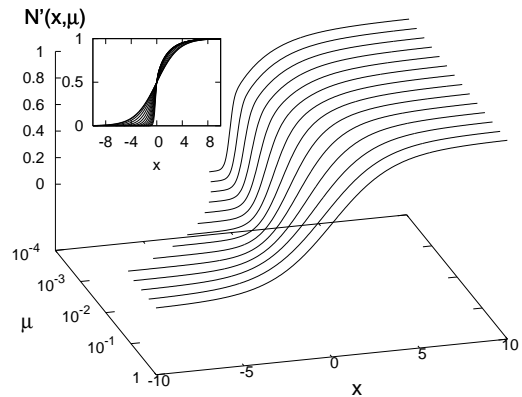,width=5.8cm,angle=-90}}
\caption
{\small Total density profiles (normalized) as a function of mass ratio $\mu$. Notice the logarithmic scale
in the $\mu$-axis. The inset shows the same profiles in a 2d plot for better comparison.}
\label{}
\end{figure}

\begin{figure}[t]
\centerline{
\psfig{file=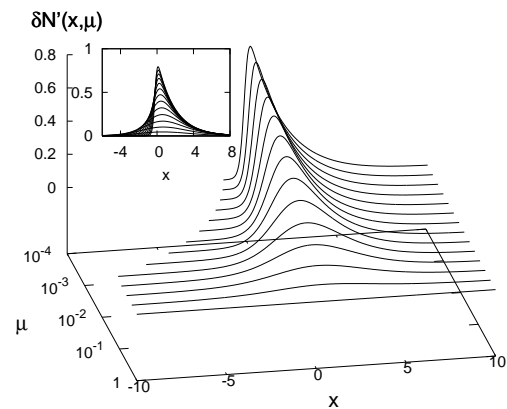,width=5.8cm,angle=-90}}
\caption
{\small Excess density profiles (normalized).}
\label{}
\end{figure}

\begin{figure}[t]
\centerline{
\psfig{file=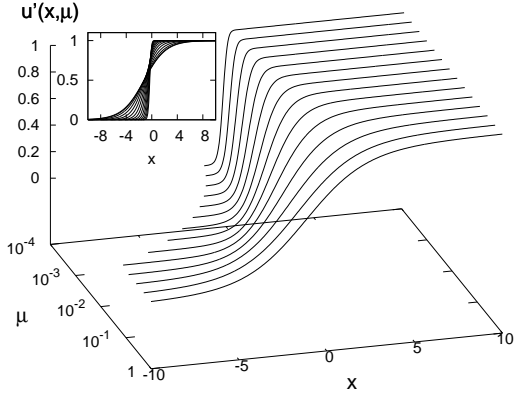,width=5.8cm,angle=-90}}
\caption
{\small Flow velocity profiles (normalized).}
\label{}
\end{figure}

We may use (\ref{decomp}) and (\ref{maxw}) to compute the local hydrodynamic currents. 
Let $J_{i,n}(x)$, $J_{i,v}(x)$ and $J_{i,e}(x)$ be the particle, momentum and energy fluxes of species
$i$ at position $x$, respectively. They can be written as,
\begin{eqnarray}
J_{i,n}(x) & \equiv & \int_{-\infty}^{+\infty}\text{d}v \ v f_i(x,v) = \tilde{n}_i\tilde{u} + \hat{n}_iu_- , \label{nflux} \\
J_{i,v}(x) & \equiv & \int_{-\infty}^{+\infty}\text{d}v \ m_iv^2 f_i(x,v) \nonumber \\
& = & m_i\big(\tilde{n}_i\tilde{u}^2 + \hat{n}_iu_-^2 \big) + \tilde{n}_i\tilde{T}_i, \label{pflux} \\
J_{i,e}(x) & \equiv & \int_{-\infty}^{+\infty}\text{d}v \ m_iv^3 f_i(x,v) \nonumber \\
& = & m_i\big(\tilde{n}_i\tilde{u}^3 + \hat{n}_iu_-^3 \big) + 3\tilde{u}\tilde{n}_i\tilde{T}_i, \label{eflux}
\end{eqnarray}
(A trivial factor $\frac{1}{2}$ has been omitted in the definition of the energy flux for convenience).
Given the constancy of the total fluxes along the system, we now write the Rankine--Hugoniot conditions as,
\begin{eqnarray}
\tilde{n}_1\tilde{u} + \hat{n}_1u_- = n_-u_-, \label{RHx1} \\
\tilde{n}_2\tilde{u} + \hat{n}_2u_- = n_-u_-, \label{RHx2} \\
m_1(\tilde{n}_1\tilde{u}^2 + \hat{n}_1u_-^2) + m_2(\tilde{n}_2\tilde{u}^2 + \hat{n}_2u_-^2) + \nonumber \\
+ \tilde{p}_1\ + \tilde{p}_2  = n_-(m_1+m_2)u_-^2, \label{RHx3} \\
m_1(\tilde{n}_1\tilde{u}^3 + \hat{n}_1u_-^3) + m_2(\tilde{n}_2\tilde{u}^3 + \hat{n}_2u_-^3) + \nonumber \\ 
+3\tilde{u}(\tilde{p}_1 + \tilde{p}_2) = n_-(m_1+m_2)u_-^3, \label{RHx4}
\end{eqnarray}
where $\tilde{p}_i =\tilde{n}_i\tilde{T}_i$ is the pressure of the regular component $\tilde{f}_i$.
Defining $\tilde{M}\equiv m_1\tilde{n}_1 + m_2\tilde{n}_2$, $\hat{M}\equiv m_1\hat{n}_1 + m_2\hat{n}_2$, 
$M_-\equiv (m_1+m_2)n_-$, and $\tilde{P}\equiv \tilde{p}_1 + \tilde{p}_2$, we obtain,
\begin{equation}
\tilde{M}\tilde{u}=\hat{\Delta}u_-, \ \tilde{M}\tilde{u}^2 + \tilde{P} = \hat{\Delta}u_-^2, \
\tilde{M}\tilde{u}^3 + 3\tilde{u}\tilde{P} = \hat{\Delta}u_-^3,
\label{RHnew}
\end{equation}
where $\hat{\Delta} \equiv M_- - \hat{M}$. Using the first relation into the second equation one has 
$\tilde{P}=\hat{\Delta}u_-(u_--\tilde{u})$, and inserting this expression into the third equation 
yields (after division by $\hat{\Delta}u_-$) $2\tilde{u}-3u_-\tilde{u}+u_-^2=0$.
This equation has solution $\tilde{u}=u_-/2$ (the second root, $\tilde{u}=u_-$, is trivial and can be neglected as 
incompatible with the boundary condition at $+\infty$). Therefore the regular flow velocity is constant along the system,
and equal to its asymptotic value at $+\infty$, $u_+$.
Using this in eqs. (\ref{RHx1})--(\ref{RHx2}) one has,
\begin{eqnarray}
\tilde{n}_i(x) & = & 2\big[n_- - \hat{n}_i(x)  \big], \quad i=1,2 \label{nregular} \\
\tilde{u}(x) & = & \frac{1}{2}u_-. \label{uregular}
\end{eqnarray}
Combination of these results with $\tilde{P}=\hat{\Delta}u_-(u_--\tilde{u})$ yields a relation for the regular pressures, namely,
\begin{equation}
\tilde{p}_1 + \tilde{p}_2 = \frac{u_-^2}{4}\big( m_1\tilde{n}_1 + m_2\tilde{n}_2\big).
\label{prel1}
\end{equation}

\begin{figure}[t]
\centerline{
\psfig{file=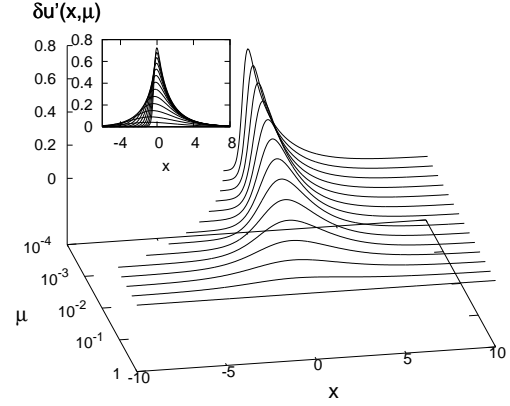,width=5.8cm,angle=-90}}
\caption
{\small Excess flow velocity profiles (normalized).}
\label{}
\end{figure}

These manipulations are intended to express all remaining flow fields in terms of the singular particle densities, $\hat{n}_i(x)$.
Equations for these singular densities can be derived
integrating with respect to $v$ eq. (\ref{singular}) for the singular component $\hat{f}_i$.\cite{sing} This yields,
\begin{equation}
u_-\frac{\text{d}\hat{n}_i(x)}{\text{d}x} = 2 \alpha_j(x) \hat{n}_i(x) \big[ \hat{n}_j(x)-n_- \big],
\label{nsingular}
\end{equation}
where,
\begin{eqnarray}
\alpha_i(x) \equiv \frac{u_-}{2} \text{erf} \Big[\sqrt{\frac{m_i}{2\tilde{T}_i(x)}}\frac{u_-}{2} \Big] 
+ \sqrt{\frac{2\tilde{T}_i(x)}{\pi m_i}}\text{exp}\Big[-\frac{m_i u_-^2}{8\tilde{T}_i(x)}\Big],
\label{alpha}
\end{eqnarray}
and where we have used results (\ref{nregular}) and (\ref{uregular}). 
To completely specify the flow fields, we need another expression relating
the regular pressures $\tilde{p}_i$ to the other fields. This relation can be derived studying the momentum and energy
transfer from species $i$ to species $j$. Multiplying eq. (\ref{eqB1}) by $m_i v$ (momentum transfer) or $m_i v^2$ 
(energy transfer) and integrating over $v$, we arrive to differential equations for $J_{i,v}(x)$ and $J_{i,e}(x)$ of the form,
\begin{eqnarray}
\frac{\text{d}J_{i,v}}{\text{d}x} & = & \Pi_i(x) \label{jvx}, \\
\frac{\text{d}J_{i,e}}{\text{d}x} & = & \Xi_i(x) \label{jex},
\end{eqnarray}
where the fluxes are defined in eqs. (\ref{pflux})--(\ref{eflux}) and,
\begin{eqnarray}
\Pi_i(x) & \equiv & \int_{-\infty}^{+\infty}\text{d}v \ m_i v \big[{\cal Q}_i(f_i,f_j)-f_i\nu(f_j)\big], \label{pidef} \\
\Xi_i(x) & \equiv & \int_{-\infty}^{+\infty}\text{d}v \ m_i v^2 \big[{\cal Q}_i(f_i,f_j)-f_i\nu(f_j)\big]. \label{xidef}
\end{eqnarray}
Since the total momentum flux, $J_{1,v}(x)+J_{2,v}(x)$, and total energy flux, $J_{1,e}(x)+J_{2,e}(x)$,  are constants along 
the line (Rankine-Hugoniot conditions),
the above differential equations, (\ref{jvx})--(\ref{jex}), express the transfer of
momentum and energy from species $i$ to species $j$ at a given point $x$. Using the expressions for $J_{i,v}(x)$ and 
$J_{i,e}(x)$ in (\ref{pflux})--(\ref{eflux}), it is easy to show that 
$3\tilde{u}J_{i,v}(x)-J_{i,e}(x)=\frac{1}{2}m_in_-u_-^3=\text{constant}$, and therefore,
\begin{equation}
3\tilde{u}\Pi_i(x)-\Xi_i(x)=0.
\label{relationpixi}
\end{equation}

This last equation will give us the desired extra relation for the regular pressures. In general, the integrals in 
(\ref{pidef})--(\ref{xidef}) cannot be performed due to the velocity--dependent collision kernel $\vert v-w\vert$ appearing
in the definition of ${\cal Q}_i$ and $\nu$, see (\ref{Qdef}) and (\ref{nudef}). In order to continue our derivation, we now 
approximate this kernel by a generic velocity--independent kernel $\sigma(x)$, much in the spirit of Maxwell molecules.\cite{Maxwell1d}
 
Using this assumption we can solve the above integrals, obtaining,
\begin{eqnarray}
\Pi_i(x) & = &  \frac{2m_im_j}{m_i+m_j}n_- u_-\sigma(x)\big(N_i-N_j \big), \label{piaprox} \\
\Xi_i(x) & =  &\frac{m_im_j}{(m_i+m_j)^2}\sigma(x) \Big\{ u_-^2 \big[ (m_i+2m_j)\tilde{n}_i\hat{n}_j  \nonumber \\
 & - & (m_j+2m_i)\hat{n}_i\tilde{n}_j \big] - 4 \big[ N_j \tilde{p}_i - N_i \tilde{p}_j \big] \Big\}, \label{xiaprox}
\end{eqnarray}
where we have defined the species total number density, $N_i(x) \equiv \tilde{n}_i(x) + \hat{n}_i(x)=2n_- - \hat{n}_i(x)$, $i=1,2$.
Within this approximation, eq. (\ref{relationpixi}) reduces to,
\begin{equation}
N_j \tilde{p}_i - N_i \tilde{p}_j = \frac{u_-}{8}(m_j-m_i)\big(\tilde{n}_i\hat{n}_j + 
\hat{n}_i\tilde{n}_j \big),
\label{prel2}
\end{equation}
which is the second relation between the regular pressures we were looking for. Together with (\ref{prel1}), this yields,
\begin{equation}
\tilde{p}_i = \frac{u_-^2\big[(m_j-m_i)(\tilde{n}_i\hat{n}_j + \hat{n}_i\tilde{n}_j) + 
2\tilde{M} N_i \big]}{8(N_i + N_j)}, \label{p1} \\
\end{equation}
where we have used the previous definition $\tilde{M} \equiv m_i\tilde{n}_i+m_j\tilde{n}_j$. 
This result, based on the Maxwell velocity kernel approximation, gives predictions for the regular pressures which are
very close $\forall x$ to the results obtained from the solution of eq. (\ref{relationpixi}) based on the numerical evaluation
of integrals (\ref{pidef})--(\ref{xidef}), thus validating our approximation. The reason behind this good agreement is that 
Eq. (\ref{prel2}) does not depend on the collision kernel $\sigma(x)$ chosen for the Maxwell-like approximation: the same 
equation is found for any kernel $\sigma(x)$ one may think of, and in particular for the (unknown) optimal kernel which better
approximates the real velocity-dependent kernel.

Recalling that $\tilde{n}_i=2(n_--\hat{n}_i)$ and $\tilde{u}=\frac{1}{2}u_-$, 
eqs. (\ref{p1}) express the regular pressures $\tilde{p}_i=\tilde{n}_i\tilde{T}_i$, and hence the regular temperatures, 
in terms of the singular densities $\hat{n}_i$. Therefore the problem of the shock wave reduces to solving the coupled, 
strongly non-linear differential equations (\ref{nsingular}) for $\hat{n}_i(x)$, $i=1,2$, since all other non-constant flow fields, 
$\tilde{n}_i$ and $\tilde{T}_i$, with $i=1,2$, can be written in terms of $\hat{n}_i$. 

\begin{figure}[t]
\centerline{
\psfig{file=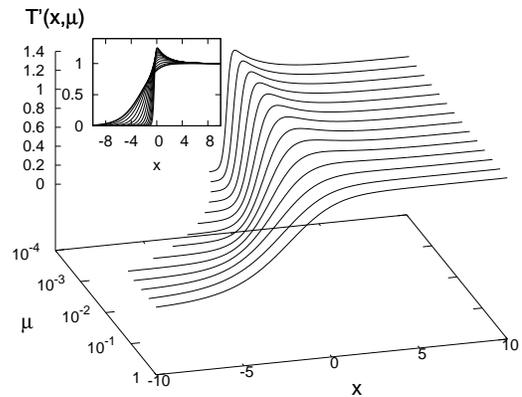,width=5.8cm,angle=-90}}
\caption
{\small Temperature profiles (normalized).}
\label{}
\end{figure}

\begin{figure}[t]
\centerline{
\psfig{file=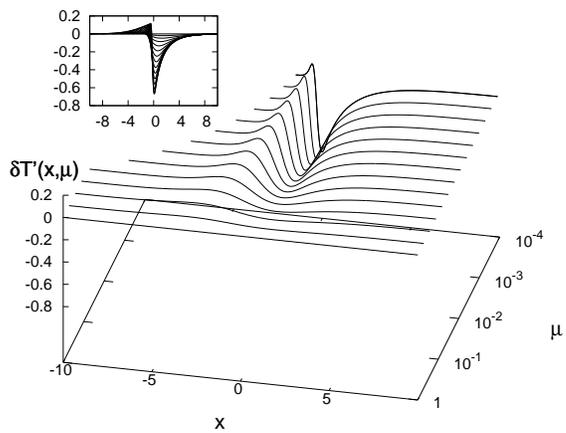,width=5.8cm,angle=-90}}
\caption
{\small Excess temperature profiles (normalized).}
\label{}
\end{figure}

\section{III. Results}

The strong non-linearities present in eqs. (\ref{nsingular}) prevent any full analytical solution to the shock wave problem,
so numerical integration must be used instead. This procedure yields the singular densities $\hat{n}_i(x)$, from which all 
hydrodynamic flow fields follow. We are particularly interested here in the total number density $N(x)$, flow velocity $u(x)$, 
and temperature $T(x)$, and their particularization to each species ($N_i(x)$, $u_i(x)$ and $T_i(x)$, $i=1,2$, 
respectively). The total number density is defined as $N(x)\equiv N_1(x) + N_2(x)$, where $N_i(x)=\tilde{n}_i(x) + \hat{n}_i(x)$
has been defined above. The total flow velocity and temperature fields are defined respectively as
$u(x) \equiv \big[\sum_i m_i N_i(x)\big]^{-1} \sum_i m_i N_i(x) u_i(x)$ and 
$T(x) \equiv N(x)^{-1} \sum_i N_i(x) T_i(x)$,\cite{Mazur} where,
\begin{eqnarray}
u_i & \equiv & N_i^{-1} \int_{-\infty}^{+\infty} \textrm{d}v \ v f_i(x,v) = \frac{n_- u_-}{N_i}, \nonumber \label{solui} \\
T_i & \equiv & N_i^{-1} \int_{-\infty}^{+\infty} \textrm{d}v \ m_i \big[v-u(x)\big]^2 f_i(x,v) \nonumber \\
 & = & \frac{\tilde{n}_i \tilde{T}_i + m_i \big[\tilde{n}_i\big(\frac{1}{2}u_--u\big)^2 + 
\hat{n}_i\big(u_--u\big)^2 \big]}{N_i}. \nonumber \label{solTi}
\end{eqnarray}

For a generic hydrodynamic flow field $\phi(x)$, we now define an associated normalized field as 
$\phi'(x)=(\phi(x)-\phi_-)/(\phi_+-\phi_-)$, where $\phi_{\pm}=\phi(x\to \pm \infty)$. This allows us to compare shock profiles
for different mass ratios $\mu \equiv m_1/m_2 \in (0,1)$. In addition, we define the normalized \emph{excess} flow fields as 
$\delta \phi' (x) \equiv \phi_2'(x) - \phi_1'(x)$ to study the species dependence of the profiles. Our results for $N'(x)$, $u'(x)$
and $T'(x)$ are given in Figs. 1, 3 and 5, respectively, while the excess fields $\delta N'(x)$, $\delta u'(x)$ and $\delta T'(x)$
are depicted in Figs. 2, 4 and 6. It should be noted that we set the coordinate origin $x=0$ so that $N'(x=0)=\frac{1}{2}$. 

A general observation is that all the profiles get steeper, and the excess fields get more localized and peaked around $x=0$, 
as $\mu$ decreases. However, there are fundamental differences for the different hydrodynamic quantities. The density flow field, 
$N'(x)$, converges toward a limiting shape as $\mu \to 0$, characterized by a sudden increase of density from $N'(x \to 0_-)=0$
to $N'(x \to 0_+)=\frac{1}{2}$, followed by a much slower
relaxation toward the asymptotic value 
$N'(x \to +\infty)=1$, see Fig. 1. On the other hand, the flow velocity profile $u(x)$ converges toward a step-like function 
localized at $x=0$, Fig. 3, while the temperature field, Fig. 5, exhibits an overshoot which sharpens and increases as $\mu \to 0$. 
In addition, all profiles are markedly asymmetrical. Fig. 7 exhibits all fields for several mass ratios.

\begin{figure}
\centerline{
\psfig{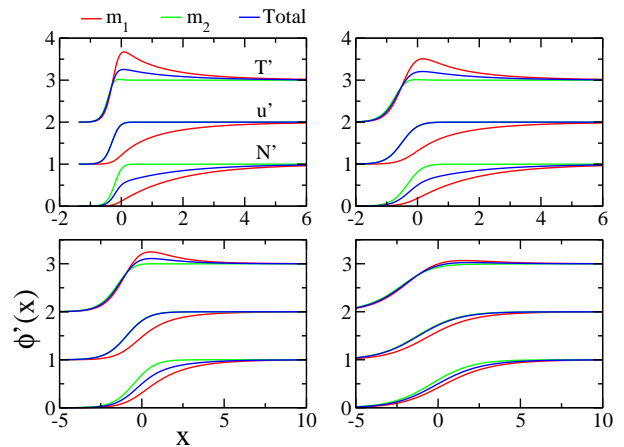}}
\caption
{\small (Color online) Density, velocity and temperature profiles for each species and the mixture. 
From left to right and top to bottom, $\mu=0.000125, 0.001, 0.016, 0.128$. Velocity 
(temperature) profiles have been shifted one (two) units in the vertical axis for visual convenience.}
\label{}
\end{figure}

The excess profiles contain the information about the relative 
local distribution of the hydrodynamic fields between the species.
In particular, $\delta N'(x)$ shows that the number density of heavy particles (type 2) around the shock layer is 
larger than that of light particles (type 1), see Fig. 2. This excess concentration of heavy particles around the shock is 
asymmetric, with most excess heavy particles lagging behind the shock, and it increases as $\mu \to 0$. On the other hand, the 
flow velocity $u_1(x)$ of light particles around the shock is larger than the velocity of heavy particles, see Fig. 4, the difference
increasing with decreasing $\mu$ (notice here that $\delta u'(x)>0$ involves $u_1(x)>u_2(x)$). 
Finally, the excess temperature field exhibits an interesting behavior, see Figs. 6 and 7. First, 
$\delta T'(x)$ is mostly negative near the shock, 
so heavy particles are cooler than light particles around the shock, despite the former are denser than the latter around $x=0$. 
More in detail, $\delta T'(x)$ exhibits an oscillation around $x=0$, meaning that heavy particles are much colder than light particles 
\emph{behind} the shock layer (the cooler the smaller $\mu$ is), while the situation is reversed \emph{in front} of the shock,
where heavy particles are now more energetic than light ones, see Fig. 6.

To better understand some of these observations, we can analyze the asymptotic behavior of eqs. (\ref{nsingular}). First,
we study how the downstream binary fluid goes toward equilibrium by letting $x \gg 0$. In this case, the 
singular densities are very small, $\hat{n}_i(x \gg 0) \ll 1$, so we can linearize eqs. (\ref{nsingular}), obtaining a simple 
decoupled system of differential equations,
\begin{equation}
u_- \frac{\textrm{d}\hat{n}_i(x)}{\textrm{d}x} = -2 \alpha_j(+\infty)n_-\hat{n}_i(x),
\label{linear}
\end{equation}
with $i=1,2$, and where,
\begin{eqnarray}
\alpha_i(+\infty) & = & \frac{u_-}{2} \Big[\textrm{erf}\big(\sqrt{\frac{m_i}{m_1+m_2}} \big) \nonumber \\
 & + & \sqrt{\frac{m_1+m_2}{\pi m_i}} \textrm{exp}\big(- \frac{m_i}{m_1+m_2}\big)  \Big]
\label{ainf}
\end{eqnarray}
is the limit of the amplitudes (\ref{alpha}) as $x \to +\infty$. 
Therefore, we find $\hat{n}_i(x) \sim \textrm{exp}[-x/\lambda_j]$ in this limit,
with $\lambda_j = u_-/[2n_- \alpha_j(+\infty)]$, so the downstream binary fluid reaches equilibrium exponentially fast,
as expected from the uncorrelated nature of Boltzmann equation.\cite{Resibois,Mazur,Harris} 
This equilibration process is characterized however by two different typical scales: $\lambda_1$ controls how heavy particles (type 2) 
relax asymptotically, while $\lambda_2$ controls the behavior of light particles (type 1). The $\mu$-dependence of both typical 
scales is depicted in Fig. 8. Here we see that $\lambda_2 > \lambda_1$ $\forall \mu \in (0,1)$. In fact, $\lambda_2$ converges to 
a constant as $\mu \to 0$, while $\lambda_1$ goes asymptotically to 0 as $\mu^{1/2}$, see eq. (\ref{ainf}). Therefore, a separation
of scales emerges for very small mass ratios $\mu$, with both fast and slow evolution scales controlling the fluid relaxation. 
All the macroscopic hydrodynamic fields $\phi(x)$ relax asymptotically as $\textrm{exp}[-x/\lambda_2]$, following the slowest
relaxation scale associated to light particles, except for the number density for heavy particles, which relax much faster, 
as $\textrm{exp}[-x/\lambda_1]$.

\begin{figure}
\centerline{
\psfig{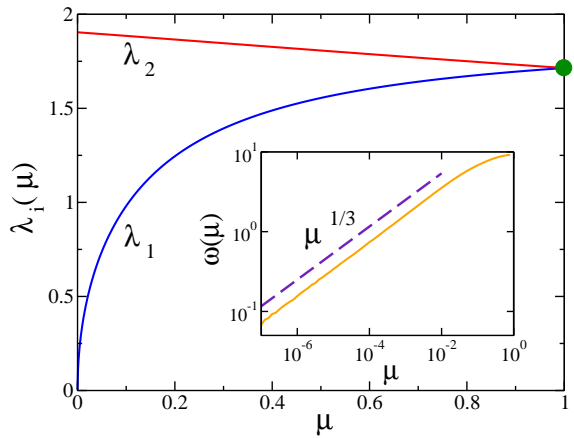}}
\caption
{\small (Color online) Main: Mass ratio dependence of the typical scales $\lambda_i$ controlling the exponential relaxation of the 
downstream binary fluid toward equilibrium. Notice that $\lambda_1$ scales as $\mu^{1/2}$ for small $\mu$. 
The solid dot signals the singular limit $\mu=1$, for which both typical scales formally diverge.
Inset: Mass ratio dependence of the shock width $\omega$ in log-log scale. The width decreases as $\omega\sim \mu^{1/3}$
for $\mu \to 0$.}
\label{}
\end{figure}

In the opposite limit, $\mu \to 1$, a singularity emerges. In fact, Boltzmann equation does not yield any useful information 
for $\mu=1$ in one dimension (1d): the gain and loss terms in the collision operator are equal in this case, so the collision 
term in Boltzmann equation is zero (see right hand side of Eqs. (\ref{eqB1})--(\ref{eqB2})). That is, collisions do not change 
the velocity distributions $f_i(x,v)$ in this limit, as expected from the fact that equal-mass particles exchange their 
velocities upon collision in 1d. Therefore there is no relaxation of hydrodynamic fields for $\mu=1$, and 
the relaxation scales $\lambda_i$ are not defined. Alternatively, this can be interpreted as a divergence of both 
$\lambda_i$. However, this is a singular behavior which emerges only for $\mu=1$. As soon as 
$\mu<1$, normal behavior is recovered. In particular, for $\mu$ slightly smaller than 1, one expects both typical scales to 
be almost equal, as observed in Fig. 8.

The separation of scales described above for $\mu \to 0$ does not only emerge for $x \to +\infty$, but characterizes the profiles 
$\forall x$. In fact, the amplitudes $\alpha_i(x)$, see eqs. (\ref{alpha}), are such that $\alpha_1(x) \gg \alpha_2(x)$ $\forall x$ as 
$\mu \to 0$. Hence, in this limit the singular density $\hat{n}_2(x)$ changes at a much faster rate than $\hat{n}_1(x)$ does,
see eqs. (\ref{nsingular}), so we can consider  $\hat{n}_1(x)$ as a constant in the scale in which $\hat{n}_2(x)$ evolves.
This effectively decouples again the differential equations (\ref{nsingular}) (though they are still highly non-linear), giving rise to 
a multi-scale relaxation process: (i) In a first step,
heavy particles very quickly relax toward their equilibrium downstream asymptotic state while light particles remain close to 
the upstream state. (ii) This is followed by a relatively much slower relaxation of light particles toward equilibrium. A good illustration
of this process can be found in the top-left panel in Fig. 7. This two-scales relaxation mechanism for $\mu \to 0$ explains the limiting 
shape described above for $N'(x)$, see Fig. 1, as well as the the positive excess density $\delta N'(x)$ in Fig. 2. In addition, it also 
explains the step-like limiting profile for the flow velocity: $u(x)$ is the weighted average of the species flow velocities $u_i(x)$,
the weigh being proportional to the species mass, so for $\mu \to 0$ $u(x)$ coincides with $u_2(x)$, which is controlled by the fastest
relaxation scale.

It seems also interesting to characterize the region where the fluid is far from equilibrium. A \emph{canonical} measure of the scale 
or typical size associated to this region is given by the shock width $\omega$. The shock width is usually defined as the inverse of 
the maximum density profile derivative, $\omega(\mu) \equiv (\frac{\textrm{d}N}{\textrm{d}x})_{max}^{-1}$.\cite{Whitham} It can
be also defined as the standard deviation of the (peaked) \emph{distribution} $A N'(x)[1-N'(x)]$, with $A$ a suitable normalization 
constant. Both definitions give equivalent results. The inset to Fig. 8 shows our results for $\omega(\mu)$. The shock width decreases
as the mass ratio decreases, scaling as $\omega(\mu)\sim \mu^{1/3}$ for $\mu \to 0$.\cite{width}
Therefore nonequilibrium effects are more localized around $x=0$ for smaller $\mu$. 
However, it is remarkable that despite the shock wave gets steeper ($\omega$ decreases) as $\mu$ decreases, the typical scale
characterizing relaxation toward equilibrium, $\lambda_2$, increases with $\mu$. Hence, although the strong nonequilibrium 
region is reduced as $\mu \to 0$, the consequent fluid evolution to equilibrium slows down.

\section{IV. Conclusions}

In this paper we have studied the response of a one-dimensional binary fluid to a strong shock wave propagating into it, 
on the basis of Boltzmann equation. This excitation drives the system far from equilibrium, therefore allowing us to 
investigate the structure of the fluid under nonequilibrium conditions and how equilibrium is regained asymptotically. 

Extending to fluid mixtures an elegant method by H. Grad,\cite{Grad,Cercignani} we have obtained the shock hydrodynamic profiles
characterizing the transition between the two different asymptotic equilibrium states. In particular, we determine the flow fields as a 
function of the mixture mass ratio $\mu=m_1/m_2\in(0,1)$. We find in general that all profiles sharpen as
$\mu \to 0$. The particle number density field converges in this limit to an asymptotic shape characterized by a sudden step-like 
increase followed by a much slower relaxation to equilibrium. The flow velocity profile converges in turn to a 
step-like function, while the temperature field exhibits a characteristic overshoot which increases with decreasing  $\mu$. 
In addition, the density (velocity) of heavy particles \emph{behind} the shock is larger (smaller) than that of light particles, the 
differences increasing as $\mu \to 0$. On the other hand, heavy particles are much cooler than light particles right behind the
shock, while the reverse situation holds in front of the shock, where heavy particles are now slightly more energetic than light ones.

In order to understand these results, we have performed an asymptotic analysis of our equations. This reveals the emergence of 
two very different typical length scales controlling relaxation as $\mu \to 0$. In this limit, the fluid evolves toward equilibrium in twos steps:
(i) First, heavy particles quickly relax toward their asymptotic downstream equilibrium state, while light particles remain close to the
upstream state; (ii) at a much slower pace, light particles come to equilibrium. In this way, light particles ultimately control the fluid's 
global equilibration. Far behind the shock layer, the hydrodynamic fields relax exponentially. Here the multi-scale relaxation mechanism 
is also apparent, with two typical exponential scales, $\lambda_i$, $i=1,2$. In particular, $\lambda_1$, associated to heavy particles,
goes to zero as $\mu^{1/2}$, while  $\lambda_2$, which controls the global relaxation to equilibrium, increases as $\mu \to 0$, 
converging to a constant. In addition, the size of the strong nonequilibrium region associated to the shock layer, as measured by the
shock width $\omega(\mu)$, decreases as $\omega(\mu)\sim \mu^{1/3}$ for $\mu \to 0$. Therefore, the fluid's strong 
nonequilibrium behavior gets more localized around $x=0$ as $\mu$ decreases, despite its asymptotic relaxation to equilibrium,
as given by $\lambda_2$, slows down with decreasing $\mu$. 

These results are specially interesting at the light of recent numerical studies of similar 1d binary 
fluids.\cite{Fourier1,Fourier2,untercio,uncuarto,Jou,Pablo} In particular, simulations show that whenever the 1d binary 
fluid is perturbed away from equilibrium, the consequent relaxation happens in such a way that light particles always tend 
to absorb more energy than heavy ones, as predicted here. In addition, light particles are 
observed to equilibrate more slowly, therefore controlling the system global relaxation. This 
supports the presence of two different relaxation scales for $\mu \to 0$, validating our theoretical results. 

The success of our Boltzmann equation approach to describe some of the structural and relaxational anomalies observed in 
the 1d diatomic fluid would suggest extending the present approach to understand its anomalous transport properties 
(see Section I)\cite{Fourier1,Fourier2,Narayan,untercio,dosquintos,Boltzmann,uncuarto}. Unfortunately, this is not possible. 
Anomalous transport in 1d is a direct consequence of the long-time power-law tails 
characterizing the decay of correlation functions in the fluid. The microscopic origin of these tails is 
associated to dynamically correlated sequences of collisions among fluid molecules, which are not described by 
Boltzmann equation.\cite{spectrum} That is, the main hypothesis on which Boltzmann equation is based, namely the 
\emph{molecular chaos} hypothesis,\cite{Resibois,Mazur,Harris} breaks correlations present in the fluid which are 
responsible of the power-law tails.

The strong correlations emerging in 1d fluids, and its role in the shock wave problem, are interesting 
issue which deserve attention.\cite{Pablo,Andreev} Further work in this direction, taking into 
account these correlations in generalized kinetic theories, would be highly desirable to 
better understand the nonequilibrium behavior of simple one-dimensional systems.

\vspace{0.3cm}

\noindent {\bf Acknowledgments:} The author thanks P.L. Garrido and P.L. Krapivsky for useful comments and discussions,
as well as the Spanish MEC for financial support.


\begin{thebibliography}{99}
\bibitem{equalmass} For $m_1=m_2$, the fluid studied in this paper reduces to 
an ideal one-dimensional gas of non-interacting particles. This system is not ergodic, and lacks any relaxation 
mechanism in velocity space capable of driving the system toward local thermodynamic equilibrium.
\bibitem{equaldens} Notice that the asymptotic number densities $n_{\pm}$ do not depend on the species, since we 
are interested here on a 1d binary fluid with \emph{equal} concentrations of light and heavy particles.
\bibitem{trivialsol} In this process we disregard a trivial solution $u_+=u_-$ which is incompatible with the boundary 
condition at $+\infty$.
\bibitem{width} It should be noticed that, due to the separation of scales as $\mu \to 0$, $\omega(\mu)$ is given in this limit 
by $u_-/[2\alpha_1(x)\hat{n}_1(x)(n_--\hat{n}_2(x))]_{max}$.
\bibitem{spectrum} More technically, Boltzmann equation for hard spheres (in any dimension) is unable to predict the long-time 
behavior of correlation functions due to the discrete character of the collision operator spectrum. In this case 
the long-time behavior is determined by the smallest non-zero eigenvalue of the collision operator, and this yields an exponential 
decay of correlations, instead of the observed power-law.\cite{Boltzmann}
\end{thebibliography}
\end{document}